\documentclass[12pt]{article}
\usepackage{amsmath}
\usepackage{graphicx}
\usepackage{enumerate}
\usepackage{natbib}
\usepackage{url} % not crucial - just used below for the URL 

\usepackage{mathrsfs}
\usepackage{amsmath,amssymb}
\usepackage{bm}
\usepackage{natbib}
\usepackage[usenames]{color}
\usepackage{amsthm}
\usepackage{arydshln}

\usepackage{multirow} 
\usepackage{enumitem}
\usepackage{caption}
\usepackage{subfig}
\usepackage{enumitem}
\usepackage{dsfont}
\usepackage{mathtools}
\usepackage{caption}
\usepackage{booktabs}

\usepackage[colorlinks,
linkcolor=red,
anchorcolor=blue,
citecolor=blue
]{hyperref}

\usepackage{mylatexstyle}
\newcommand{\blind}{1}

\usepackage{xcolor}

\ifdefined\final
\usepackage[disable]{todonotes}
\else
\usepackage[textsize=tiny]{todonotes}
\fi
\allowdisplaybreaks

\begin{document}

\def\spacingset#1{\renewcommand{\baselinestretch}%
{#1}\small\normalsize} \spacingset{1}

%%%%%%%%%%%%%%%%%%%%%%%%%%%%%%%%%%%%%%%%%%%%%%%%%%%%%%%%%%%%%%%%%%%%%%%%%%%%%%

\if1\blind
{
  \title{\bf From Authors to Reviewers: Leveraging Rankings to Improve Peer Review\thanks{This paper is a discussion of the 2025 JASA discussion paper by \citet{su2025ICML}.}}
  \author{Weichen Wang\textsuperscript{1} and Chengchun Shi\textsuperscript{2}\\
  \\
   \textsuperscript{1}Faculty of Business and Economics, University of Hong Kong\\
   \\
   \textsuperscript{2}Department of Statistics, London School of Economics and Political Science}
   \date{\empty}
  \maketitle
} \fi

\if0\blind
{
  \bigskip
  \bigskip
  \bigskip
  \begin{center}
    {\LARGE\bf Discussion: Can Reviewers’ Opinions of the Ranking of Reviewed Papers Assist Peer Review in Machine Learning?}
\end{center}
  \medskip
} \fi

%\bigskip
%\begin{abstract}
%\end{abstract}

%\noindent%
%{\it Keywords:}  
%\vfill

%\newpage
\spacingset{1.9} % DON'T change the spacing!

\section{Introduction}

We would like to congratulate the authors on conducting a comprehensive and insightful empirical investigation of the 2023 ICML ranking data. The review quality of machine learning (ML) conferences has become a big concern in recent years, due to the rapidly growing number of submitted manuscripts. In particular, in 2025, NeurIPS received roughly 30k submissions, while other top conferences -- such as ICML, ICLR, CVPR, and AAAI -- each received over 10k. Managing such extraordinarily large submission pools is extremely challenging, which leads to noisy and low-quality reviews. This decline in review quality has hampered the selection of high-quality papers and, ultimately, the advancement of ML as a field.  
%we hope to, on the one hand, reconsider the design of the review system that more accountably evaluates the quality of submitted papers, and on the other hand, improve the data analysis of ranking papers beyond naively averaging reviewer scores. 

To address this issue, two main approaches have been proposed: (i) redesigning the review system to incentivize reviewers into consistently providing high-quality reviews, or to identify low-quality reviews with input from authors \citep[see e.g.,][]{kim2025position}; and (ii) applying data-driven methods to calibrate review scores. 
The authors adopted the second approach and made a valuable contribution by incorporating author's ranking information via isotonic regression to improve review scores. Specifically, they collaborated with the 2023 ICML conference to conduct an experiment in which authors submitting more than one paper were asked to rank their own submissions. Their empirical results show that using these author-provided rankings yields much smaller mean squared and absolute errors than those of the raw review scores \citep{su2025ICML}. Theoretically,  \cite{su2021you, su2022truthful} proved that under certain conditions, the author's optimal strategy to maximize their utility is to honestly report the ranking of their own papers. 

Although the empirical study clearly demonstrates the advantage of leveraging authors' provided paper rankings, we see two potential limitations to this approach. The first is that, if the isotonic mechanism were implemented in practice, some authors might inevitably attempt to game the system in an adversarial manner. The theoretical results of \citet{su2021you, su2022truthful}, which show that authors should truthfully report their paper rankings, were established under strong assumptions that ranking is done only once for a single conference. In reality, this assumption is clearly violated: major conferences are held annually, and authors could iteratively experiment with different ``ranking strategies'' over time. 
The authors also commented that ``it is crucial to note that the rankings were provided under the condition that they would not influence decision-making processes. Authors might behave strategically if their rankings were used in decision-making'' \citep{su2025ICML}.  We appreciate their cautiousness in implementing the proposed methodology. 

Second, there is a fairness concern regarding authors with differing numbers of submissions. The isotonic mechanism only applies to authors with more than one submission, and the evaluation tends to improve with more papers submitted. This may result in more active machine learning researchers receiving better reviews, while newer researchers experience noisier evaluations.

In this discussion, we propose an alternative approach that leverages ranking information from reviewers rather than authors. This approach avoids the aforementioned two limitations and can also be integrated with the isotonic mechanism. Our main idea is to explore deeply how reviewers generate their scores. Conventionally, it is assumed that reviewers review each assigned paper independently, with scores centered around the true quality of the submission plus some random noise, modeled as e.g.,  Gaussian\footnote{Note that the true score is not truly continuous like a Gaussian but rather categorical; here, the Gaussian assumption is used for simplification.}. However, we observe two departures from this assumption.  
First, the independence assumption can be violated. When a reviewer is assigned multiple papers, they are likely to compare these papers, causing their scores to follow a multivariate Gaussian distribution conditioned on their ranking. Specifically, the reviewer may first score each assigned paper and then adjust the scores via the isotonic mechanism based on their ranking preferences to ensure that the paper they like best receives the highest score. Second, review scores can be biased depending on each reviewer's experience and scoring habits. However, this bias is typically constant across all papers assigned to the same reviewer and therefore does not affect their rankings. Based on these two observations, we propose the following model for the score generating process (SGP). 

\begin{assumption}[SGP]\label{assump:SGP}
\label{ass:assump1}
Each submitted paper has true quality $\theta^*_i, i=1,\dots,n$. Each submission is assigned to a few reviewers and each reviewer reviews a few submissions. Let $N_r\subseteq \{1,\cdots,n\}$ denote the papers assigned to the $r$th reviewer. Their review scores are generated following three steps: 
%Reviewer $r$ reviewed papers $i \in N_r$ with $|N_r| = R$. 
%The reviewing scores are then generated following 3 steps. 
\begin{enumerate}
\item %With an unknown bias $b_r$ and variance $\sigma_r^2$, 
The reviewer first generates raw scores independently for each paper $\tilde S_{r,i} \sim \mathcal{N}(\theta^*_i + b_r, \sigma_r)$, for each $i \in N_r$, where $b_r$ and $\sigma_r^2$ denotes the bias and variance terms specific to the $r$th reviewer.
\item After reading all assigned papers, the reviewer ranks them according to the Plackett-Luce model. That is, if $N_r = \{j_1, j_2, \dots, j_R\}$, we have
\begin{equation*}
\mathbb{P}(j_1 \succ j_2 \succ \dots \succ j_R) = \frac{\exp(\theta^*_{j_1})}{\sum_{u=1}^R \exp(\theta^*_{j_u})} \cdot \frac{\exp(\theta^*_{j_2})}{\sum_{u=2}^R \exp(\theta^*_{j_u})} \cdot \cdots \cdot \frac{\exp(\theta^*_{j_{R-1}})}{\sum_{u=R-1}^R \exp(\theta^*_{j_u})}\,,
\end{equation*}
where $i \succ j$ indicates that paper $i$ is preferred over paper $j$, and the reviewer samples a ranking according to this probability distribution.  
\item The final scores $\{S_{r,i}\}$ are generated following the isotonic regression to make sure the raw scores $\{\widetilde{S}_{r,i}\}$ generated in Step 1 satisfy the sampled ranking in Step 2. Specifically, $\{S_{r,i}\}_{i\in N_r}$ are obtained by solving
\[
\min_{S_{r,i}, i\in N_r} \sum_{i\in N_r} (S_{r,i} - \tilde S_{r,i})^2  \quad \text{s.t.} \quad  S_{r,j_1} \ge S_{r,j_2} \ge \dots \ge S_{r,j_R},
\]
assuming the ranking is given by $j_1\succ j_2 \cdots \succ j_R$. 
\end{enumerate}
\end{assumption}

%We will leave the study of inference for $\theta_i^*, b_r, \sigma_r$ under this SGP for future research. However, we hope to provide some numerical results 
We notice that the model in generating the raw score in Step 1 shares similar spirits with the two-way fixed-effects model in the panel data literature \citep[see e.g.,][]{imai2021use}, where $\widetilde{S}_{r,i}$ is determined additively by the paper's quality and the reviewer's bias, without an interaction term. 
In the next section, we conduct a simulation study showing that if this SGP holds, leveraging ranking information from reviewers can lead to more accurate evaluation of each paper's quality $\theta_i^*$ in the presence of reviewers' biases $\{b_r\}_r$. In contrast, if the reviewer is unbiased (i.e. $b_r = 0$), then simply averaging the review score $\tilde S_{r,i}$ over $r$ yields the most accurate paper scores possible. However, given the limited number of highly experienced and perfectly impartial reviewers, as well as the extraordinarily large number of paper submissions, it is reasonable to assume that reviewers' biases are prevalent in practice.

\section{Simulation Studies}

In this section, we simulate review data based on the SGP described in Assumption \ref{assump:SGP} and design a setting that closely mimics the 2023 ICML conference submissions. Our results show that (i) incorporating ranking information from reviewers can significantly improve the evaluation of each paper’s quality, often outperforming the use of ranking information from authors alone; and (ii) combining ranking information from \textit{both} reviewers and authors yields the most accurate evaluation of submitted papers in most scenarios.

\begin{figure}[t!]
    \centering
    \subfloat[\#authors per paper]{\includegraphics[width=0.33\textwidth]{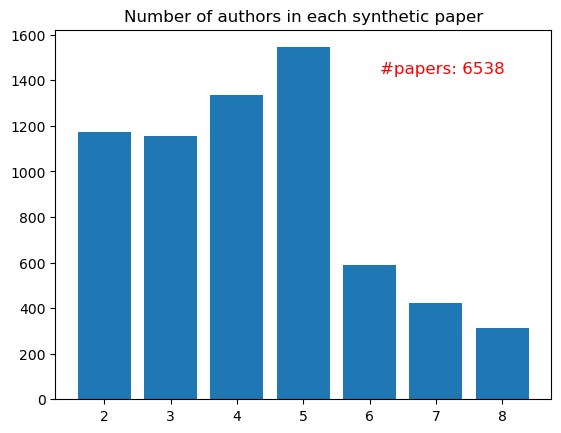}\label{fig:1a}}
    \subfloat[\#papers per author]{\includegraphics[width=0.33\textwidth]{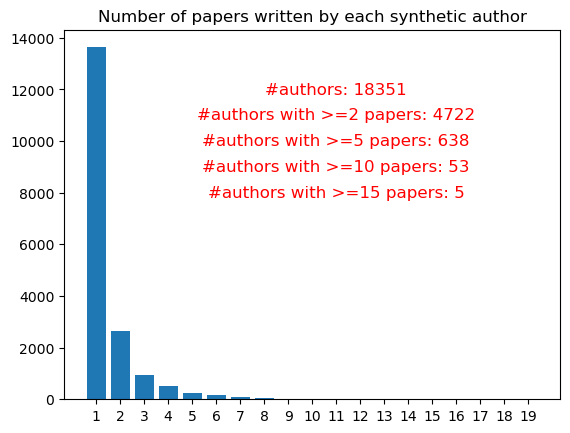}\label{fig:1b}}
    \subfloat[\#reviewers per paper]{\includegraphics[width=0.33\textwidth]{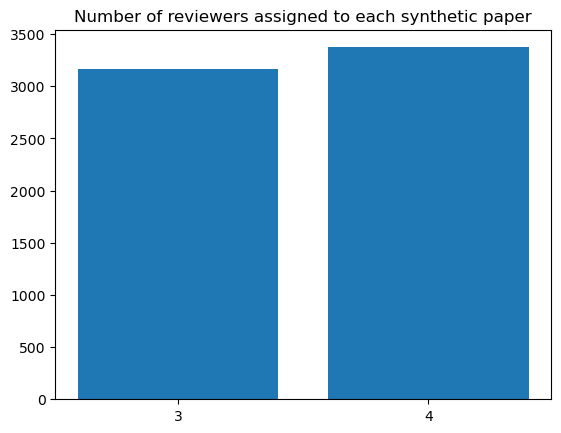}\label{fig:1c}}

    \subfloat[\#paper per reviewer]{\includegraphics[width=0.33\textwidth]{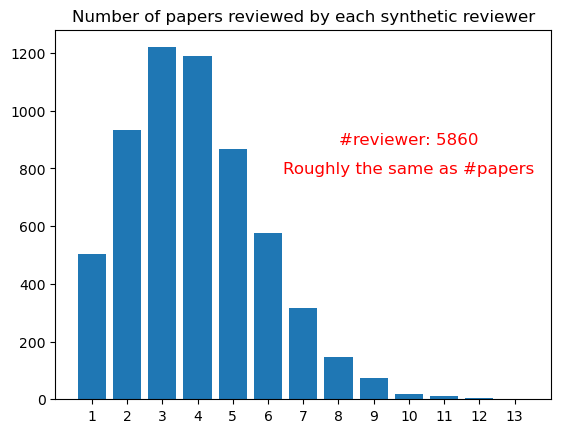}\label{fig:1d}}
    \subfloat[Density of true scores]{\includegraphics[width=0.33\textwidth]{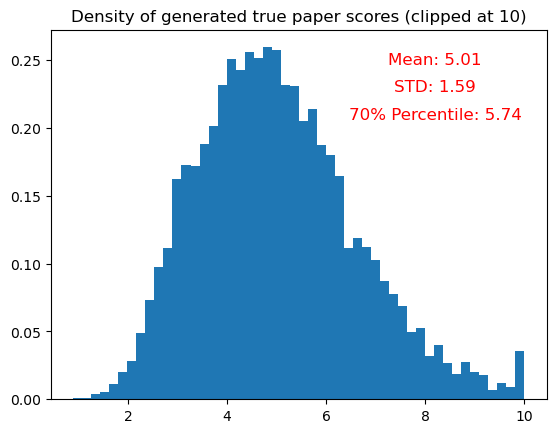}\label{fig:1e}}
    \subfloat[Histogram of reviewer scores]{\includegraphics[width=0.33\textwidth]{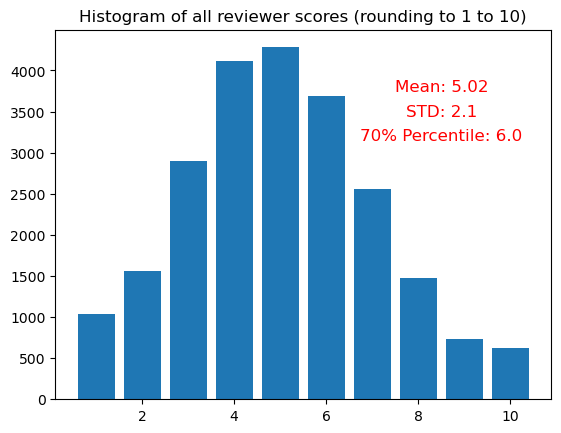}\label{fig:1f}}
\caption{Summary statistics and distributions of the synthetic data.}\label{fig:1}
\end{figure}

First, we use the summary statistics of the 2023 ICML conference submissions reported by the authors to calibrate our simulation model, generating paper submissions, authors, reviewers and true paper scores. We simulate two groups of authors: pure ML researchers who frequently publish papers in ML conferences and interdisciplinary ML researchers who only occasionally do so. %These groups 
The first group represents authors who are likely to submit multiple papers to the conference. 
%The first group is used to match the tail behavior of the distribution of the number of submissions of each author. 
According to \cite{su2025ICML}, the total number of submissions is $6538$ and the total number of authors is $18535$. 
%Assuming on average one paper has $3$ to $4$ authors, each author is expected to submit $1.06$ to $1.41$ papers. 
The number of authors with at least $2, 5, 10, 15$ submissions is reported to be $4505, 508, 74, 26$ out of $18,535$ authors, correspondingly $24.3\%, 2.74\%, 0.40\%$ and $0.15\%$, respectively. The second group corresponds to the majority of authors ($75.5\%$) submitting only one paper and is represented by the leftmost bar in the histogram in Figure \ref{fig:1b}. %of single submission in the histogram in Figure \ref{fig:1b} is better interpreted by researchers who are new to the machine learning field and do not submit more than one paper.   
For each simulation, we create $6,538$ synthetic submissions. Figures \ref{fig:1a}, \ref{fig:1b}, \ref{fig:1c}, \ref{fig:1d}, \ref{fig:1e} present respectively, the number of authors in each paper, the number of papers each author writes, the number of reviewers assigned to each paper, the number of papers each reviewer reviews and the density of true paper scores respectively. 
%and the histogram of generated reviewer scores following Assumption \ref{ass:assump1}. In the generation of the reviewer scores, we consider the following five cases:
Next, following Assumption \ref{ass:assump1}, we consider the following five cases to generate the reviewer scores: 
\begin{itemize}
\item Base case: $b_r \sim Unif(-2,2), \sigma_r \sim Gamma(1,1)$;
\item No-Bias case: $b_r = 0, \sigma_r \sim Gamma(1,1)$;
\item No-Variance case: $b_r \sim Unif(-2,2), \sigma_r = 0$;
\item Big-Bias case: $b_r \sim Unif(-3,3), \sigma_r \sim Gamma(1,1)$;
\item Big-Variance case: $b_r \sim Unif(-2,2), \sigma_r \sim Gamma(1,1.5)$.
\end{itemize}

\noindent Figure \ref{fig:1f} visualizes the histogram of our simulated review scores in the base case. The detailed code to generate papers, authors and reviewers can be found at \url{https://github.com/weichwang/ReviewerPaperRanking}.

Second, we consider four different methods for estimating the true score $\theta_i^*$ and compare their root mean square errors (RMSEs). The first method naively takes average over reviewers of the reviewer scores for each paper, which aligns with the current practice in most ML conferences. The benchmark method is expected to work well without reviewers' biases but can suffer from large RMSEs with moderate to large biases. The second method is our proposed approach that combines ranking information from reviewers. The main idea is to first use a variant of the method proposed by \cite{fan2025spectral} (outlined in the next section) to estimate a full ranking of all submissions using only the reviewers’ partial rankings of the papers they review, and then apply isotonic regression to calibrate the average review scores. The third approach is the one proposed by \cite{su2025ICML}, which leverages the true rankings provided by authors and applies the isotonic regression for calibration. The last method combines the second and third approaches: we adjust the average scores using rankings from both the reviewers and the authors. %Note that the reviewer rankings are based on the PL model, therefore contains randomness, whereas the author rankings are assumed to follow the true paper scores.

\begin{table}[t]
\centering{
\caption{Comparison of 4 Methods in RMSE across 10 repetitions}
\label{tab1}
\begin{tabular}{l|ccccc}
\hline
RMSE                                                      & \multicolumn{1}{c}{Base}                                 & \multicolumn{1}{c}{No-Bias}                        & \multicolumn{1}{c}{No-Var}                   & \multicolumn{1}{c}{Big-Bias}                      & \multicolumn{1}{c}{Big-Var}                  \\
\hline
1. Average Scores (Benchmark) & \begin{tabular}[c]{@{}c@{}}0.850 \\ (0.013)\end{tabular} & \begin{tabular}[c]{@{}c@{}}0.589 \\ (0.010)\end{tabular} & \begin{tabular}[c]{@{}c@{}}0.693\\ (0.006)\end{tabular} & \begin{tabular}[c]{@{}c@{}}1.077\\ (0.016)\end{tabular} & \begin{tabular}[c]{@{}c@{}}0.948\\ (0.013)\end{tabular} \\
\hline
2. Use Rankings from Reviewers                            & \begin{tabular}[c]{@{}c@{}}0.752 \\ (0.014)\end{tabular} & \begin{tabular}[c]{@{}c@{}}0.614\\ (0.006)\end{tabular}  & \begin{tabular}[c]{@{}c@{}}0.598\\ (0.005)\end{tabular} & \begin{tabular}[c]{@{}c@{}}0.888\\ (0.016)\end{tabular} & \begin{tabular}[c]{@{}c@{}}0.836\\ (0.012)\end{tabular} \\
\hline
3. Use Rankings from Authors                        & \begin{tabular}[c]{@{}c@{}}0.787 \\ (0.011)\end{tabular} & \begin{tabular}[c]{@{}c@{}}{\bf 0.555}\\ (0.008)\end{tabular}  & \begin{tabular}[c]{@{}c@{}}0.650\\ (0.006)\end{tabular} & \begin{tabular}[c]{@{}c@{}}0.984\\ (0.014)\end{tabular} & \begin{tabular}[c]{@{}c@{}}0.872\\ (0.011)\end{tabular} \\
\hline
4. Combine 2 and 3                                        & \begin{tabular}[c]{@{}c@{}}{\bf 0.709} \\ (0.012)\end{tabular} & \begin{tabular}[c]{@{}c@{}}0.585\\ (0.005)\end{tabular}  & \begin{tabular}[c]{@{}c@{}}{\bf 0.571}\\ (0.006)\end{tabular} & \begin{tabular}[c]{@{}c@{}}{\bf 0.829}\\ (0.013)\end{tabular} & \begin{tabular}[c]{@{}c@{}}{\bf 0.782}\\ (0.010)\end{tabular} \\
\hline                           
\end{tabular}}
\end{table}

Finally, Table \ref{tab1} reports the RMSEs of the estimated true scores for the 6538 submissions,  obtained by the four methods under the five cases. The true paper scores, authors, and reviewers are generated only once, while the review scores are generated 10 times. We then report the average RMSE across these 10 simulations, with standard deviations shown in parentheses. Although we only simulate the data for 10 times, the RMSEs are highly stable with small standard deviations, due to the large number of submissions. It is clear from Table \ref{tab1} that except for the No-Bias case, rankings from the reviewers can improve the RMSE quite significantly. The improvement is particularly pronounced in the Big-Bias case. Additionally, author rankings always improve the RMSE regardless of whether reviewers' biases are present or not. This is expected since author rankings are generated to reflect the true paper scores perfectly. Interestingly, except for the No-Bias case, the improvement from author rankings is slightly smaller than that achieved using reviewer rankings, even though reviewer rankings are generated according to the PL model and therefore contain randomness. Lastly, combining both reviewer and author rankings generally attains the best performance.

\section{Methodology}
%We first extract small-set partial rankings from each reviewer for those papers reviewed. %Note that these rankings are robust towards any level of parallel score shift. 
%We first briefly review the spectral ranking method proposed by \cite{fan2025spectral}. 
Given a small set of partial rankings from each reviewer for the papers they review, we can employ the spectral ranking method by  \cite{fan2025spectral} to estimate a complete ranking of all papers. Specifically, \cite{fan2025spectral} construct a Markov chain in which transitions are determined by which paper ``wins'' over another and  show that the stationary distribution of this transition matrix $P$ encodes information about each $\theta_i^*$. Specifically, let $\pi$ denote the population-level stationary distribution; it can be shown that the probability of selecting the $i$th paper satisfies $\pi_i \propto \exp(\theta_i^*)$. Consequently, we can estimate $\hat\theta_i$ from the empirical stationary distribution $\hat\pi$. For further details, we refer readers to \cite{fan2025spectral}.
%Finally we can adjust the benchmark average scores for all papers by this inferred ranking, again following the isotonic regression. 

In our implementation, we obtain the full ranking of papers from the reviewer scores using a hierarchical ranking approach, rather than the original spectral ranking method of \cite{fan2025spectral}. Given the limited sample nature of paper review process, one paper is only compared to a very small number of other papers, through 3 or 4 reviewers. If one paper has a relatively large $\theta_i^*$, it is very likely that this paper will never lose in any comparisons with other papers. In fact, out of the $6,538$ simulated papers, about $1,000$ papers are always selected as the top preferred paper by each reviewer. If we were to use spectral ranking, the transition probability $P_{ij}$ essentially measures how likely paper $i$ will lose to paper $j$. As a result, for these 1,000 papers, the transition probabilities to other papers would be zero, leaving no data to differentiate them. Consequently, spectral ranking would fail to rank these papers solely based on reviewer rankings.

Our hierarchical ranking solves this issue with the following procedure. We start from all papers and use the corresponding transition matrix $P$ to identify the top-tier paper group $G_1$, which contains papers that never lose to any other papers. Then we remove $G_1$ from the current paper list, and reconstruct the transition matrix with the remaining papers. The new transition matrix will give us the second-tier group $G_2$, which contains papers that never lose within the remaining papers. We continue to remove $G_2$ from the current paper list and the procedure stops when we cannot identify any more papers that never lose to other papers. The remaining papers will be the last-tier group, say $G_{K+1}$ if we run the hierarchical ranking for $K$ times. Based on this procedure, it can be inferred that papers in $G_k$ should be ranked higher than papers in $G_{k+1}$ for $k=1,\dots, K$. Within each tier group, since we do not have effective rankings, we use the average review scores to rank papers. This procedure produces a full ranking over all papers, which is then used to apply isotonic regression and estimate the paper quality scores.

The final adjusted score is defined as $1/2$ times the isotonic optimized scores plus $1/2$ times the original average review scores. This averaging accounts for the fact that reviewer rankings are not perfectly accurate, effectively regularizing the isotonic optimized scores toward the original average review scores. We do not optimize the tuning parameter $1/2$ in this illustration, as the results are stable with respect to this choice.

Finally, we remark that our newly proposed method is intended only for illustration. It is a heuristic proposal without any theoretical support at this stage. However, it showcases that combining ranking information from reviewers (possibly together with rankings from authors), has large potential to improve the review quality. Because this approach relies solely on reviewer rankings, it is inherently fair regardless of the number of submissions per author. Another advantage is that it does not require to implement a new review system; simply changing how review scores are processed can already yield significant improvements. 
This direction is worth of further investigation in the future. We would like to hear any thoughts from the authors along this line. Congratulations again on this comprehensive and thought-provoking work.

\bibliographystyle{agsm}

\bibliography{ref}

\end{document}